\newif\ifieee
\ifieee
	\documentclass{IEEEcsmag}
\else
	\documentclass{article}
	\usepackage{color}
	\usepackage{graphicx}
\fi
\usepackage[colorlinks,urlcolor=blue,linkcolor=blue,citecolor=blue]{hyperref}
\definecolor{darkgreen}{rgb}{0.0, 0.4, 0.4}

\newcommand{\mytt}[1]{{\texttt{#1}}}

\newcommand{\hangpara}{
 \setlength{\parindent}{0cm} 
 \hangindent=0.4cm 
}

\ifieee
\jvol{XX}
\jnum{XX}
\paper{XX}
\jmonth{XX/XX}
\jname{Computer}
\pubyear{2022}
\fi

\begin{document}


\title{CUF-Links: Continuous and Ubiquitous FAIRness Linkages for reproducible research}

\ifieee
	\author{Ian Foster}
	\affil{Argonne National Laboratory \& The University of Chicago}
	\author{Carl Kesselman}
	\affil{University of Southern California}
\else
	\author{Ian Foster and Carl Kesselman}
	\date{}
\fi

\markboth{}{Paper title}

\begin{abstract}
Despite much creative work on methods and tools, reproducibility---the ability to repeat the computational steps used to obtain a research result---remains elusive. One reason for these difficulties is that extant tools for capturing research processes do not align well with the rich working practices of scientists. We advocate here for simple mechanisms that can be integrated easily with current work practices to capture basic information about every data product consumed or produced in a project. We argue that by thus extending the scope of findable, accessible, interoperable, and reusable (FAIR) data in both time and space to enable the creation of a continuous chain of continuous and ubiquitous FAIRness linkages (CUF-Links) from inputs to outputs, such mechanisms can provide a strong foundation for documenting the provenance linkages that are essential to reproducible research.  We give examples of  mechanisms that can achieve these goals, and review how they have been applied in practice.

\end{abstract}

\maketitle

\section{INTRODUCTION}

The ability to repeat the research used to produce a particular finding or discovery has long been viewed as an important element of the scientific method~\cite{Popper,national2019reproducibility}.
In order to permit such repetition, means must exist for documenting and communicating the steps involved in that research.
Thus, scientific papers typically include
a natural language description
of how a result was obtained. We also see the emergence of systems
for recording and sharing experimental protocols~\cite{teytelman2016protocols}.

With advances in the ability to collect source data, and the increased use of computational tools, the task of 
documenting scientific methods has expanded to include 
recording not only the \textit{methods used to generate data}, with a view to enabling studies aimed at collecting new data to answer the same scientific question, but also the subsequent \textit{computational steps applied to data}, once generated, to obtain a result. 
A 2019 NASEM report uses the terms \textit{reproducibility} to refer to the latter concern,
which it defines as
``obtaining consistent results using the same input data, computational steps, methods, code, and conditions of analysis''~\cite{national2019reproducibility}.

One might think that reproducibility should be easy to achieve: given that computation is performed by (usually deterministic) digital computers under the direction of software, researchers should need only to record the precise sequence of logical operations applied to input data to generate outputs.
Yet despite much creative work on methods and tools, 
reproducibility remains an elusive goal.
A frequent reason, in our experience, is the loss of linkages between inputs and outputs in multi-step computational processes, particularly when different steps are performed at different times, by different people, with different tools, or in different computational environments---as, for example, when a dataset is copied from one machine to another, or transferred to a collaborator.

We argue that the solution to such problems is not, in general, to restrict analyses to occur only in closed environments (e.g., Jupyter notebooks) which, while recording linkages automatically, inevitably constrain as well as empower. 
Taking a socio-technical perspective~\cite{baxter2011socio}, we instead seek to identify minimal mechanisms that can be used to enhance existing research processes in ways that improve both reproducibility and usability without introducing significant new costs.
Our overarching goal here is ensure that \textit{all} data and programs are Findable, Accessible, Interoperable, and Reusable (FAIR), regardless of \textit{when} and \textit{where} they are produced---a discipline that we refer to as Continuous and Ubiquitous FAIRness (CUF)---with the goal of capturing and maintaining linkages among these artifacts: what we refer to as CUF-Links.  
With such linkages in place, reproducibility becomes easier.

In the sections that follow, we review the challenges inherent in documenting processes for research projects that may engage many people over extended periods;
present six principles that we have found useful in achieving CUF-Links goals;
describe methods and tools that we have used to establish and maintain CUF-Links;
and present an example of the application of these methods and tools.

\section{CONTINUITY AND UBIQUITY}

A typical research project involves repeated hypothesis generation, experiment, and data analysis activities. These activities often extend over a considerable period of time; involve multiple people; employ different experimental and computational methods and apparatus; produce and consume many intermediate datasets; and involve many choices by programs and people.

At one or more points in a project, a researcher typically wants to publish results. They review the data collected to date, select those relevant to the planned publication, perhaps perform some additional analyses, and prepare figures and tables. 
At this point, they are challenged by the NASEM report to ``\textit{convey clear, specific, and complete information about any computational methods and data products that support [those] results}''---information that should encompass the elements listed in \autoref{tab:reqs}.

\begin{table*}
\caption{Reproducibility requires that results be accompanied by clear, specific, and complete information on~\cite{national2019reproducibility}:}\label{tab:reqs}

\centering
\small
\setlength{\tabcolsep}{12pt}
\ifieee
	\begin{tabular*}{\textwidth}{@{\hspace{1ex}} p{5.8cm}  p{3.5cm}  p{4cm} }
\else
	\begin{tabular*}{\textwidth}{@{\hspace{1ex}} p{4.4cm}  p{2.8cm}  p{3.1cm} }
\fi
\multicolumn{1}{c}{\textbf{Data}} & \multicolumn{1}{c}{\textbf{Methods}} & \multicolumn{1}{c}{\textbf{Environment}} \\
\textsf{The input data used in the study, either in extension (e.g., text file or binary) or intension (e.g., a script to generate the data), as well as intermediate results and output data for steps that are nondeterministic and cannot be reproduced in principle.} & 
\textsf{A detailed description of the study methods (ideally in executable form) together with its computational steps and associated parameters.} &
\textsf{Information about the computational environment where the study was originally executed, such as operating system, hardware architecture, and library dependencies.}
\end{tabular*}
\end{table*}

Anecdotal reports, and our own experience, suggest that this transition process is painful and error-prone, requiring hunting down data that may be distributed across notebooks, laptops, cloud storage, and local storage systems, often with inadequate metadata, and attempting to reconstruct, from ad hoc assemblages of directory structures, file names, spreadsheets, and text files, what is going into the paper. And despite best intentions, these steps are often unsuccessful, as noted by Tedersoo et al.~\cite{tedersoo2021data}, who in a survey of researchers found ``not enough time to search'' and ``couldn’t find the data'' reported as the top two reasons for not sharing data.

A major reason for these difficulties is that, as noted above, the data, study methods, and computational environments that a paper's author(s) seek to document have 
considerable temporal and spatial extent. 
Time is a source of entropy and an enemy of memory; space multiplies the number of entities that must be considered when seeking to capture data, methods, and environment.
Furthermore, the computational steps that generate a particular research output were likely performed by busy, distracted, and fallible people, employing a range of computational methods and tools, and using unreliable devices (human memory, notes, filenames, conversations) to transfer information among steps.
It is little wonder if researchers cannot recall the data, study methods, and computational environments used to produce a result. 
And without these details, it is often impossible to define what data were used in an analysis, what version of an analysis code was used to create a graph, and so on.

We thus conclude that any approach to reproduciblity
that focuses only on the moment of data publication is doomed to failure.
If we do not record 
the identity, location, and nature of the data (and associated methods) used to generate each intermediate data item at the time it was created, we are unlikely to be to recall those steps later.
In other words, the solution to poor reproducibility is more FAIRness: in fact, \textbf{continuous and ubiquitous FAIRness} that encompasses the full temporal and spatial extent of a research project.

\begin{figure}[b]
    \centering
    \includegraphics[width=\columnwidth]{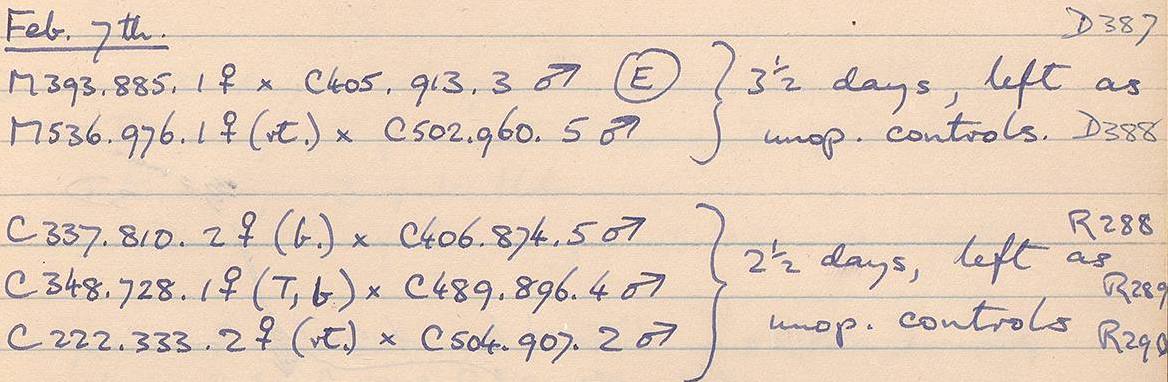}
    \caption{One day's entry in development biologist Anne McLaren's notebook on embryo transfer experiments in mice, 1950s. (British Library, shelfmark Add MS 83844; copyright estate of Anne McLaren).}
    \label{fig:mclaren}
\end{figure}

This goal of documenting every step followed in a research project is far from new. Indeed, a nineteenth century scientist would find it familiar, and point to their laboratory notebook, in which they 
``wr[o]te with enough detail and clarity that another scientist could pick up the notebook at some time in the future, repeat the work based on the written descriptions, and make the same observations that were originally recorded"~\cite{kanare1985writing}.
Indeed, notebooks (whether analog or digital) are still in use today.
However, while recording details in this way could be easy when a scientist labored for a day to perform a single experiment (see, for example, \autoref{fig:mclaren}), it becomes harder as the use of digital technologies, automated data collection methods, and computational processes multiplies the number of experiments, data products, and methods employed. 

Or does it?  Digital technologies certainly increase the amount of work to be recorded, but computerized methods also offer opportunities to improve on the past. If an operation is performed by a computer, then both its details and the data that it produces can, in principle, be recorded automatically, in detail, and without risk of error. 

The difficulty, in many cases, is no longer that of recording details but of maintaining connections between entities so that we can determine at a later date that \textit{this result} (e.g., the contents of a file \textit{F}) was generated with \textit{this method} (e.g., \textit{M}) running on \textit{this computer} (e.g., \textit{C}).
In other words, we need to assign identifiers (e.g., \textit{IF}, \textit{IM}, \textit{IC}) to entities; link each identifier with the entity identified (its referent): e.g., \textit{IM--M}; ensure that identifier-referent relationships are preserved; and use identifiers to document entity-entity relationships, so that we can record simply that ``file \textit{IF} was produced by method \textit{IM} running on computer \textit{IC}.''
This information, if appropriately collected and maintained, constitutes the continuous and ubiquitous FAIRness linkages (``CUF-Links") needed to connect research outputs back to resource inputs.

\section{PRINCIPLES}

How are we to generate the information needed to identify CUF-Links?
One approach, frequently advocated 
but less often applied successfully, is to standardize on a specific programming technology (e.g., a scripting language, workflow tool, virtual machine, container) that maintains such relationships automatically.
However, such proposals invariably fail to address the full complexity of modern research projects, in which data are produced, manipulated, and analyzed in different ways, at different times, and by different people.

A socio-technical perspective leads us to seek, instead, simple mechanisms for creating CUF-Links that can be composed easily with the rich tools and working styles that are common in research practice. 
We introduce these mechanisms here in terms of six motivating principles.

\paragraph{P1: Identify everything}
FAIR principles emphasize the importance of assigning unique and persistent identifiers to published data~\cite{force112014guiding}. Yet such identifiers are typically generated only for published data, and only at the time of data publication. Continuous and ubiquitous FAIRness demands that we have such identifiers for any artifact. To this end, we need identifiers that can be created quickly, easily, and cheaply; can be created for many different data, at varying levels of resolution (from a single file to larger collections); use a checksum to record a binding to a specific sequence of bits, so that modifications can be detected; and are portable, so that they remain understandable when taken from one context to another (e.g., when shared with a collaborator). 
The latter requirement means, in general, that identifiers need to be globally unique and supported by a mechanism to resolve them to physical addresses. 

As we explain below, we have found that these needs can be addressed by using \textit{Minids}~\cite{chard2016ll}: identifiers that are sufficiently simple to make creation and use trivial, while still having enough substance to make data easily findable and accessible. Minids can be created in the thousands (or millions), resolved quickly, and upgraded to digital object identifiers (DOIs) if data are published. 

\paragraph{P2: Hold metadata close}
FAIR principles also emphasize the need to associate metadata with data. 
The commonly used approach of \textit{embedding} selected metadata values in directory and file names ensures that metadata will not be lost when data are shared or transferred, but limits the variety of metadata that can recorded.
On the other hand, decoupling data and metadata by storing the latter in a database (perhaps indexed by file name or experiment identifier), while potentially far more expressive, is labor intensive to set up, and complicates data sharing in that any data sharing operation requires extraction of database records---records that may not be understandable by the recipient. 

A promising answer, in our experience, is in effect a compromise between these two approaches of embedding and decoupling, namely to locate metadata \textit{alongside} data. 
This approach is supported, for example, by the BagIt~\cite{kunze2018bagit} specification, and the associated BDBag and Bagit/Research Object (RO) profiles, for which we provide more details below.

\paragraph{P3: Describe things, consistently}

The FAIR principles state that \textit{metadata and data should be richly described with a plurality of accurate and relevant attributes}.
BDBag mechanisms provide a convenient mechanism for packaging such descriptions with data.

A key question is what descriptive terms to use.
In an established field, community defined terms are often available and should be used whenever possible to describe data as they are generated. 
For yet greater precision, term/ontology lookup services exist to facilitate locating unique identifiers for terms. For example, the EMBL-EBI Ontology Lookup Service~\cite{malone2015new} allows us to determine that the term \textit{DNase-Seq} (used in the first of two examples presented below) has identifier \textit{NCIT:C106052} in the National Cancer Institute thesaurus. 

Particularly in exploratory research, no appropriate term may exist. One should then define a ``local'' term, to be reused within a collaboration, with the understanding that over time, this term may be replaced by a more broadly recognized term~\cite{ribes2005learning}.
What should be avoided is the use of ``free text'' for describing data generated during an investigation, as this will just push back the inevitable task of applying consistent descriptions, perhaps to a last-minute scurry to prepare data to publication---increasing the potential for missed data and misunderstanding. 

Consistent terminology use requires a mix of both tools and best practices.
We illustrate this point with an anecdote concerning a collaboration that sought to crystallize (a trial-and-error process) a number of complex proteins and determine their structure using  X-ray crystallography. 
The informatics member of the team created a database to hold details about the various experiments (a \textit{good tool}), but left the column describing experiment status as free text (\textit{bad practice}). Unconstrained by tools or convention, researchers entered a variety of values (e.g., \textit{complete}, \textit{Complete}, \textit{completed}, \textit{Completed}, \textit{completed contaminated}, \textit{inprgress}, \textit{inprogress}, \textit{In Progress}). This lack of common terminology hindered determination of which experiments had completed and ultimately resulted in replicated experiments. 
Fixing this situation required refactoring the database to have a proper term table, creating foreign keys, mapping all existing data to those foreign keys, and rewriting the offending metadata values. We speak to evolving terms and metadata structure in our discussion of \textbf{P4}.

\paragraph{P4: Anticipate change}
Research 
often requires the invention of new concepts and terminologies that may over time, as new connections are made, be further specialized and/or merge with other ideas and nomenclature. Thus any system for organizing metadata, however informal, should facilitate both the definition of conventions (\textbf{P3}) and the evolution of those conventions over time.  

Change can occur for many reasons. We may need a new term to describe a new observation,  discover a new type of entity, or need (because of new understanding) to record more detail than previously. From a technology perspective, change may simply involve adding a new term to a term table, or alternatively may require evolution of the data model (entities and relationships) for our research domain. In the database community, a general solution to the \textit{schema evolution} program is challenging~\cite{stonebraker2016database}. For the practicing researcher, less comprehensive solutions can be helpful. For example, it is often good practice to organize metadata, from the start, to include well-defined points for augmenting term sets, and to allow descriptions of core entities to be extended by using standard techniques such as extension tables in a database. Tools designed to enable researchers (or informatics-savvy members of a research team) to modify data representations incrementally can help maintain metadata descriptions to be representative of the current structure of a research project~\cite{schuler2020towards}.

\paragraph{P5: Never repeat}
Identifying everything and holding metadata close 
can go a long way towards realizing our goal of continuous and ubiquitous FAIRness by documenting \textit{what} was manipulated at each step in a scientific project.
The \textit{how}, on the other hand, often involves tasks performed by humans---creating files, running programs, renaming files, extracting metadata, etc.---that are particularly amenable to being forgotten or performed inconsistently.
A simple practice that can greatly enhance reproducibility is to capture any task that is to be performed more than once in a program.
Shell scripts,  workflow systems, virtual machines, containers, and notebooks are all used for this purpose.
Regardless of the technology used, the important thing is that repeated steps are captured in executable form, and that this program is itself accessible via an identifier, immutable, and associated with metadata. 

Automation programs that operate on big data and computation often involve remote data access, data movement, and computation steps. In such cases, an underlying data fabric, such as that provided by Globus~\cite{chard2016globus}, can be invaluable as an enabler of programmatic access to secure, reliable, and efficient remote data and computation.
Both examples below leverage such a data fabric.

\paragraph{P6: Play it again---and again}
Adherence to the first five principles increases the likelihood that any particular data product can be connected, via an unbroken chain of CUF-Links, back to its origins.
Yet a single missing link---a bad identifier, missing metadata---can disrupt the chain.
Thus, it is important to verify, frequently and repeatedly, the ability to resolve provenance queries.
That may sound hard, but in fact it is entirely consistent with the practice of continuous integration, which we should already be following to validate the correctness of other aspects of our research processes~\cite{fowler2006continuous}.

\vspace{1ex}

These six principles complement, rather than replace, the overarching vade mecum that \textit{good science requires good software}.
The need to follow proven practices such as automating repeated actions, using version control to track the evolution of software systems, documenting assumptions, and generating unit tests are, we hope, taken as a given~\cite{sandve2013ten}.

\section{METHODOLOGY}

The methods by which the above principles are integrated into daily practice will vary depending on the tools being used by the research study, the types of data generated, the nature and size of the research team, and other factors. 
Nevertheless, we can define a general methodology that may be applied to a wide variety of use cases, focusing on computational analysis steps.
We express the following steps in terms of a singleton input dataset, analysis code, etc., but they are easily expanded to multiples.

\begin{enumerate}

    \item 
    Prior to starting an investigation, identify any existing controlled vocabularies that may be applicable. Create a \textbf{data dictionary}---which can simply be a document, spreadsheet, Python dictionary, etc.---listing the terms that you plan to use (\textbf{P3}).
    You will consult that data dictionary throughput the investigation.

    \item \label{step:input}
    Prior to running any analysis, ensure that the input dataset that you are to analyze has a globally unique \textbf{identifier} (\textbf{P1}). Use the identifier to retrieve the input dataset. If the input is not in an existing \textbf{collection} (e.g., a BDBag) create one to represent it (\textbf{P2}).

    \item \label{step:repo}
    Check out the \textbf{analysis code} from its code repository (e.g., GitHub). Record the version number or hash code for the version that you are using (\textbf{P1}). If using a single-file program, such as a spreadsheet, obtain its identifier. 
    
    \item \label{step:code}
    If the \textbf{analysis workflow} to be run is not already in the code checked out in \#\ref{step:repo}, assemble it in executable form (e.g., shell script, Jupyter notebook) and record it in a repository (\textbf{P5}). Execute the workflow.
    
    \item \label{step:id}
    Use a tool such as BDBag to create a \textbf{detailed manifest} for the analysis output and any associated metadata, to include the identifier for the code from \#\ref{step:code}.
    Create an identifier for the output dataset.
    
    \item \label{step:output}
    Place the output dataset description from \#\ref{step:id}, along with any output files referenced by that description, in a \textbf{shared repository}, and obtain an identifier for this output (\textbf{P1}).
    
    \item 
    Verify that steps \#\ref{step:input}--\ref{step:output} can be replicated, and correct if not (\textbf{P6}).
    
    \item 
    Share the \textbf{output identifier} from \#\ref{step:id}. (Do not reference individual files!) Encourage a collaborator to report their experiences with using the shared identifier to replay the analysis. 
\end{enumerate}

These steps should appear trivial: their purpose is not to change the way that research is performed, but to codify the discipline that will ensure that an unbroken chain of CUF-Links extends from original input to published output.
They can, and should, all be automatable. 
The methodology builds on best-practice software engineering methods, and strikes a balance between needs for reproducibility and low costs. Empirical observations in ongoing studies~\cite{madduri2019reproducible} show it can work in practice.

\section{BDBAGS and MINIDS}

The six principles recommend best practices, not specific technologies: they can be realized in a myriad of ways. 
Nevertheless, we describe here three technologies that we have found effective in establishing CUF-Links.
Informed by the requirements in \autoref{fig:bdbag_reqs}, these technologies comprise two main components.

The \textbf{BDBag} provides a mechanism for \emph{defining} a dataset and its contents by enumerating its elements, regardless of their location (Enumeration, Fixity, Distribution).
A BDBag is a BagIt package that conforms to BDBag and Bagit/Research Object (RO) profiles.
The BagIt specification itself defines a directory structure and required and optional files within that structure, as well as a set of methods to serialize the directory into an established archive format, such as ZIP. 
Data files in a bag are located in a payload directory named \mytt{data}, within which files may be hierarchically organized using subdirectories, while metadata files are located in a directory named \mytt{metadata}.
The BDBag and Bagit/RO profiles further specify the use of a \mytt{fetch.txt} file, require serialization,
and specify what manifests must be provided, in order to support the use of bags for big, distributed data.
RO conventions~\cite{bechhofer2013linked} allow contents of a dataset to be described for purposes of reuse; they provide a means for \emph{characterizing} a dataset and its contents with arbitrary levels of detail, regardless of data location (Description).)

\begin{figure}
\small
\ifieee
	\framebox{\parbox{0.92\columnwidth}{
\else
	\framebox{\parbox{\columnwidth}{
\fi
\hangpara
{\bf 1. Enumeration}: The dataset's elements must be explicitly enumerated, so that subsequent additions or deletions can be detected.
(Thus not, for example, ``the contents of directory D.'')

\vspace{0.5ex}
\hangpara
{\bf 2. Fixity}: A robust way of verifying that we have the intended versions of dataset contents,
so that data consumers can detect errors in data transmission or
modifications to data elements.

\vspace{0.5ex}
\hangpara
{\bf 3. Description}: Interoperable methods for tracking the attributes (metadata) and origins (provenance) of dataset contents. 

\vspace{0.5ex}
\hangpara
{\bf 4. Identification}:  A reliable and concise way of referring to datasets
for purposes of collaboration, publication, and citation.

\vspace{0.5ex}
\hangpara
{\bf 5. Distribution}: A dataset can contain elements from more than one location.

\vspace{0.5ex}
\hangpara
{\bf 6. Simplicity}:
Methods should not 
impose significant user overhead or require that complex software be deployed on researcher computers.
}}
    \caption{Six requirements for tools to support the creation and exchange of complex, big, data collections (\emph{datasets}\/) made up of many directories and files (\emph{elements}\/). Adapted from Chard et al.~\cite{chard2016ll}.}
    \label{fig:bdbag_reqs}
\end{figure}

We use \autoref{fig:bdbaglayout} to illustrate the approach. 
This shows a bag, unimaginatively named \mytt{mybag}, that contains 
two data files, \mytt{file1} and \mytt{file2}, and a single metadata file, \mytt{annotations.txt}. These user-supplied components are located, unchanged, in \mytt{data} and \mytt{metadata} directories, respectively, within a directory named by the bag name. 
Alongside these unmodified data and metadata files are the following supporting files, created and maintained by BDBag tools, that those tools use to maintain and verify the integrity of the bag's contents:

\begin{itemize}
    \item 
\mytt{bag-info.txt}, \mytt{bagit.txt}: These two files provide metadata about the bag, and a BagIt version declaration, respectively.

\item
\mytt{manifest-\emph{alg}.txt}: 
File(s) that make the contents of the bag explicit by listing all files in the payload \mytt{data} directory along with a checksum for each file. Alternative checksum algorithms can be used, with the algorithm indicated in the manifest file name, e.g., \mytt{manifest-md5.txt} or \mytt{manifest-sha256.txt}. Software can then simply examine the manifest file to detect missing or corrupted files.

\item
\mytt{tagmanifest-\emph{alg}.txt}: 
File(s) that serve the same purpose as a data manifest file, but for the \mytt{metadata} directory. 

\item
\mytt{fetch.txt}: Provides, for any file listed in the manifest but not contained in the payload directory, a line (URL, LENGTH, FILENAME) to indicate the missing file's name and a URL from which it may be obtained. 
\item

\mytt{metadata/manifest.json}: Lists the name, media type, and semantic type of each bag resource, with JSON-LD used to link metadata with existing ontologies and vocabularies. It can also provide per-resource attribution, provenance, and annotations. 
\end{itemize}

A set of software tools, the \texttt{bdbag} utilities (\mytt{github.com/fair-research/bdbag}) support various operations on BDBags, such as their creation, validation (verifying checksums), and localization (i.e., retrieving files listed in \mytt{fetch.txt}). 

The \emph{holey bags} supported by the \mytt{fetch.txt} file represents an important element of BDBag technology. It allows a bag to provide a compact and unambigious definition of a distributed data collection in which large elements---perhaps too big to transfer explicitly---are located in cloud or enterprise storage. Having received or downloaded a holey bag, a researcher uses a single command (\mytt{bdbag --resolve-fetch all}) to localize it. 
A Globus URL can be used to indicate that a file can be fetched via the Globus high-performance, reliable transfer service~\cite{chard2016globus}.

The \textbf{Minid} (\textbf{min}imal \textbf{id}entifier) provides a method for uniquely \emph{identifying} a dataset and, if desired, its constituent elements, regardless of their location (Identification, Fixity).
Minids are designed to be lightweight and easy to create, via either command line tools or programmatic interfaces that can be called from notebooks, scripts, or other tools that a researcher might use. A Minid includes minimal metadata---author, creation time, data name, one or more URLs to the underlying data, and a checksum to validate data fixity---and is \textit{resolvable}, i.e., it can be dereferenced to the underlying data object. The Handle system is used to resolve minids in a scalable and robust way.

\begin{figure}
\small
\begin{tabbing}
aa\=aa\=aa\=aaaaaaaaaaaaaaaaaaaaa\=aa\=aa\=\kill
\texttt{\footnotesize mybag/} \>\>\>\> Top level name\\
\> \texttt{{\footnotesize bag-info.txt}} \>\>\>  Metadata for the bag\\
\> \texttt{\footnotesize bagit.txt} \>\>\>  BagIt version and encoding \\
\>  \texttt{{\textbf\footnotesize data/}} \>\>\>  The bag's contents:\\
\>\>  \textcolor{darkgreen}{\textbf{\texttt{\footnotesize file1}}} \>\>\>A first user file \\
\>\>  \textcolor{darkgreen}{\texttt{\footnotesize \textbf{file2}}}\>\>\>  A second user file \\
\>  \texttt{{\footnotesize fetch.txt}} \>\>\>  How to fetch missing elements \\
\>  \texttt{\footnotesize manifest-md5.txt} \>\>\> MD5 checksums for data files\\
\>  \texttt{\footnotesize metadata/} \>\>\>  Accompanying metadata\\
\>\>  \texttt{\textcolor{blue}{\textbf{\footnotesize manifest.json}}}\>\> \> RO metadata as JSON-LD\\
\>\>  \texttt{\textcolor{blue}{\textbf{\footnotesize annotations.txt}}}\>\> \> User annotations\\
\>  \texttt{{\footnotesize tagmanifest-md5.txt}} \>\>\>  MD5 checksum for tags
\end{tabbing}
\caption{An example BDBag. Data are contained in the \textbf{\texttt{data}} directory and metadata in the \textbf{\texttt{metadata}} directory. 
\label{fig:bdbaglayout}}
\end{figure}

\begin{figure*}[t]
    \centering
    \includegraphics[width=\textwidth]{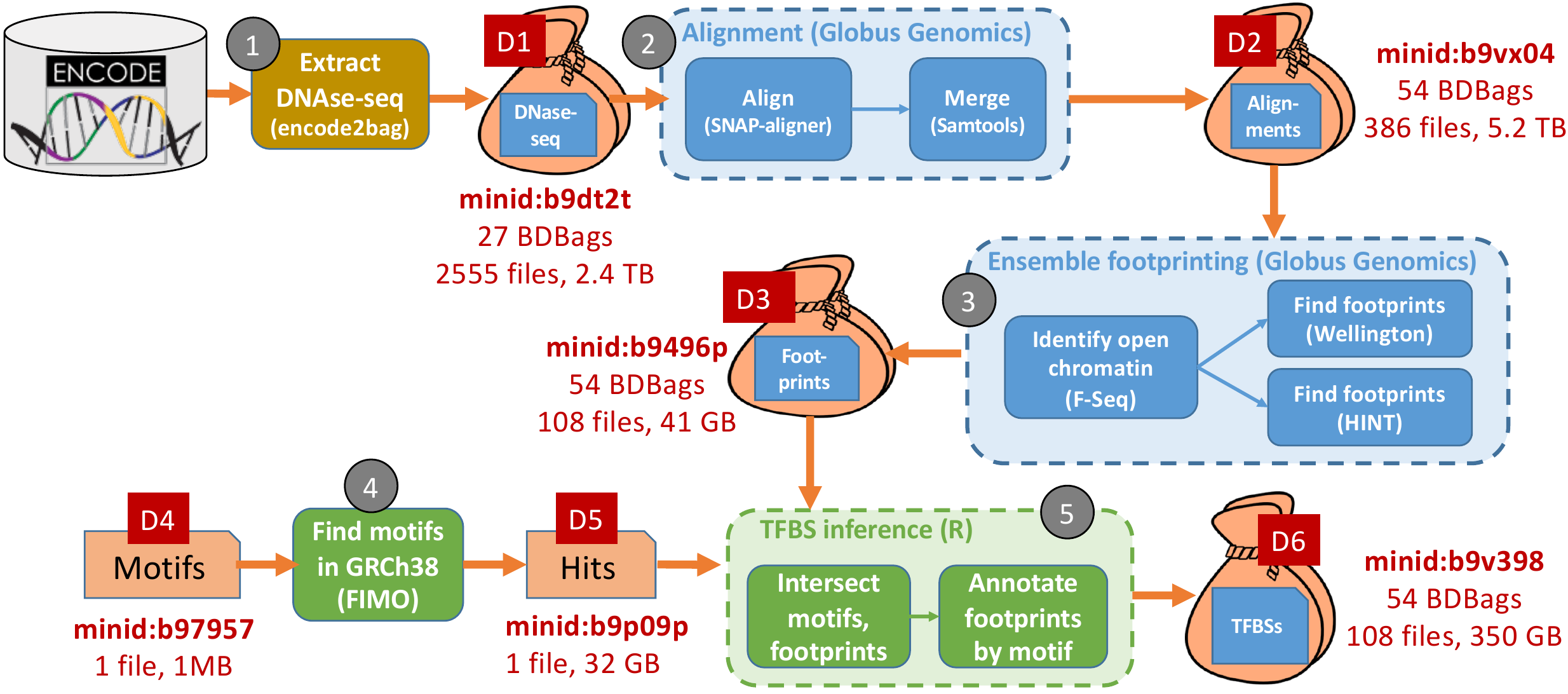}
    \caption{A high-level view of the transcription factor binding sites (TFBS) identification workflow. The six principal datasets are labeled D1--D6 (D1--D3 and D6 involve multiple BDBags, as multiple tissues are analyzed), each accessible via a Minid resolvable at \mytt{identifiers.org}, and the five computational phases are labeled 1--5: see \cite{madduri2019reproducible} for links to code.}
    \label{fig:bdds}
\end{figure*}

\section{DISCUSSION}

The Minid and BDBag mechanisms for defining, naming, and describing datasets allow us to exchange, create, and consume datasets in ways that enable continuous and ubiquitous FAIRness.  
They are neither a necessary nor sufficient solution to reproducibility: it is quite feasible to implement our principles with other technologies, as long as they are used continuously and ubiquitously---or, on the other hand, to use them without achieving reproducibility.
That said, an attractive feature of these technologies is that 
they can easily be incorporated into the fabric of even large research projects.
Because they are implemented via simple command line programs and libraries, and do not require changes to data formats or layouts,
it is straightforward, for example, to modify a program that performs a set of simulations to assign Minids for results and to assemble results into BDBags (which place metadata \textit{alongside} data, but do not modify data), and to adapt a script used to analyze simulation results  to consume BDBags and propagate identifiers.

\section{TWO SCIENCE EXAMPLES}

We use two examples to illustrate how CUF-Links methods and the tools just described can be used in practice.

The first involves a multi-step analysis used to create an atlas of putative transcription factor binding sites from terabytes of ENCODE DNase I hypersensitive sites sequencing data. As illustrated in \autoref{fig:bdds} and described in detail elsewhere~\cite{madduri2019reproducible}, this analysis involves involves multiple applications, substantial computing, and numerous large datasets.
CUF-Links methods allow the authors to record each major intermediate dataset (D1--D5) as well as the final output (D6) via a Minid. 
(The recording of intermediate datasets is important due to the cost of regenerating them.)
We note, however, that this publication refers to the software components of the analysis only by GitHub URLs, rather than specific commits, a deficiency that could allow code to change without notice. 

The second example is from a multi-year neuroscience investigation that involved the creation of a new technique for observing the location of synapses in a living organism, the development of a novel new behavioral training technique for performing unconditioned stimulus response training in larval zebrafish, and the application of these new techniques to characterize, for the first time, how synaptic structure in the brain change with learning~\cite{synapse}. 

The experiments conducted in this investigation were complex, involved many different data types, were highly collaborative, and evolved rapidly over time, often on a weekly or even daily basis. Project participants quickly realized that they would become overwhelmed by this combination of rapid iteration, complex experimental design and highly collaborative research.  Consequently, they eagerly adopted a CUF-Links based methodology implemented via Deriva~\cite{bugacov2017experiences}, a platform for scientific asset management that streamlines the integration of CUF-Links principles into daily scientific work. 

Every piece of data generated during this six-year investigation was obtained and managed following the principles outlined above.  Care was taken to ensure that the integration of CUF-Links methods did not significantly burden the researchers, so that they could integrate these principles into their daily practice. The team found that CUF-Links methods provided significant and direct benefits, including the following:

\begin{itemize}
    \item The fact that all data, including bad data, were retained along with descriptive metadata allowed the team to detect a malfunction in a microscope that was subsequently confirmed and fixed by the vendor.
    \item After developing a superior algorithm for assessing learning by the zebrafish, they could quickly locate and reprocess all behavioral observations to include the new assessments.
    \item When preparing for publication, organizing data and creating DOIs for  figures took only minutes.
    \item When responding to reviewer comments, existing experimental data could be located quickly and easily that was then analyzed further to address reviewer concerns.
    \item Updating figures and associated data to address reviewer comments took just minutes.
    \item The 
    figures and supplemental data provided complete transparency into the experiments, which all reviewers remarked on positively.
\end{itemize}

Data are organized into collections that can be exported as BDBags which uniquely identify all the elements of the collection and associated metadata. Upon export, a Minid (e.g., \mytt{minid:fPTs86M7VTyb}) is assigned to the bag to allow for citation; given such a Minid, a researcher can locate and download the bag to other computational nodes, Jupyter notebooks, etc., for further exploration. 

\begin{figure*}[t]
    \centering
    \includegraphics[width=\textwidth]{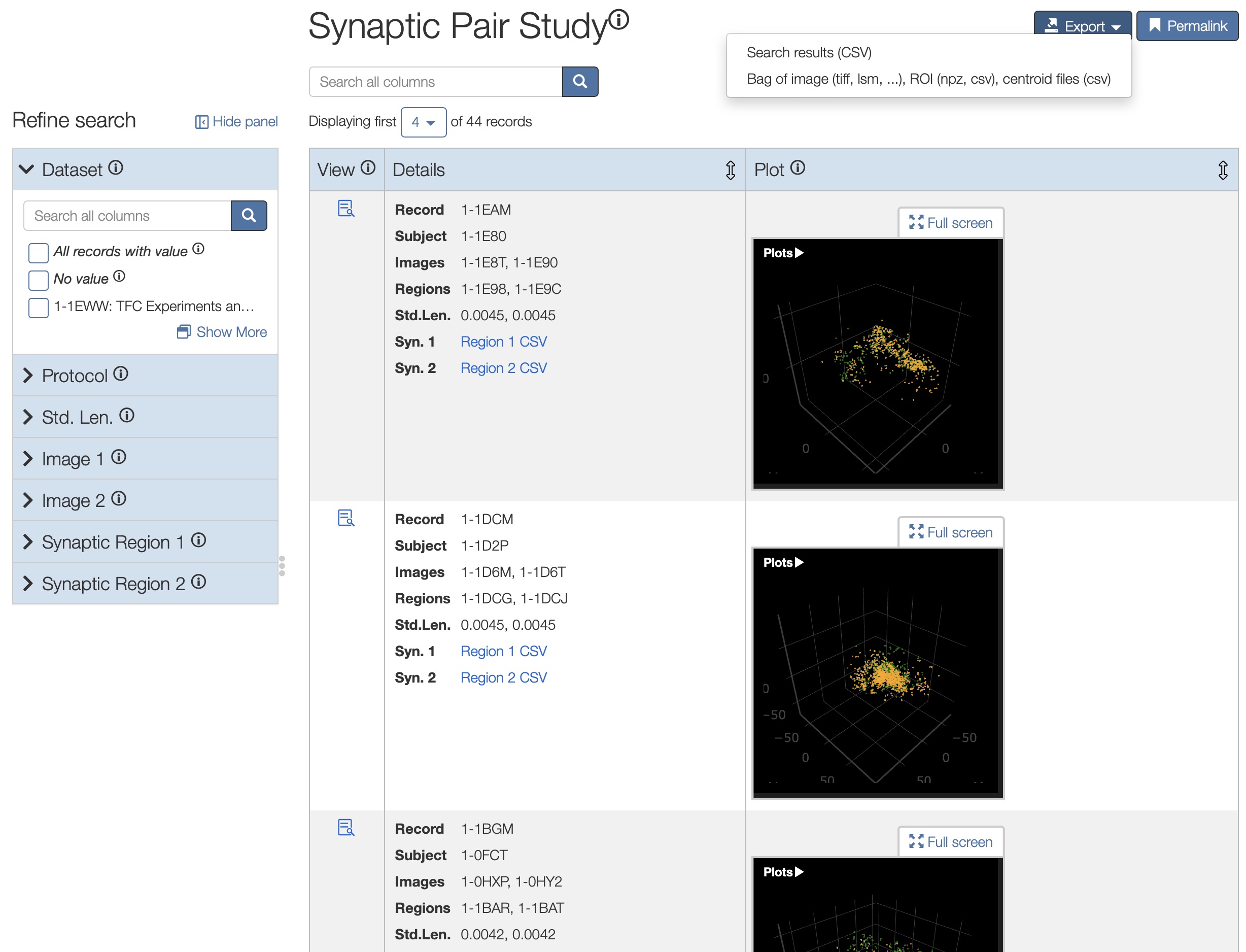}
    \caption{Screenshot of repository interface used for Synapse experiments. Facets on the left, based on the metadata model, facilitate discovery. More detailed metadata are shown in the center panel, along with interactive displays of underlying datafiles. The open ``Export'' menu on the top right shows option for extracting data as a BDBag.}
    \label{fig:synapse}
\end{figure*}

\section{CONCLUSIONS}

We have argued that the key to reproducibility, particularly in larger, collaborative projects, is to adopt tools and practices aimed at \textit{establishing and maintaining unbroken provenance links from inputs to outputs}.
We argued further that socio-technical concerns require methods by which these  continuous and ubiquitous FAIRness linkages (CUF-Links) can be created with minimal impact on the work practices of scientists. These considerations lead us first to advocate for six overarching principles and then to explain how these principles can be implemented by integrating simple mechanisms (we advocate here for BDBags and Minids) into applications and tools.  

We have focused in this article on \textit{reproducibility}, in which the computational steps applied to  data are repeated, rather than \textit{replicability}, in which new data are collected to answer the same question~\cite{national2019reproducibility}.
The distinctions between these two aspects of the research process are becoming less well-defined as,
increasingly, experimental apparatus are controlled by software, research protocols are represented in digital form, computational simulations yield valuable data, human choices feed into data analysis, and computational processes are used to determine which data to collect next. Thus CUF-Links must increasingly encompass data collection processes as well as data analysis.

\section{ACKNOWLEDGMENTS}

\ifieee
We thank the editors of this special issue, Michael Heroux, Manish Parashar, and Victoria Stodden, for encouraging us to organize and document our thoughts in this area; our collaborators on the various tools and applications referenced here for all that they have taught us; and the anonymous reviewers for their thoughtful comments.
\fi
This material was based in part upon work supported by the U.S. Department of Energy, Office of Science, under contract DE-AC02-06CH11357; by the National Institutes of Health under grants U24-CA209996 and U01-DE028729.

\ifieee
\section{ABOUT THE AUTHORS}

\textbf{Ian Foster} is the Arthur Holly Compton Distinguished Service Professor of Computer Science at the University of Chicago, and Distinguished Fellow, Senior Scientist, and Director of Data Science and Learning Division at Argonne National Laboratory. He is a Fellow of the IEEE.

\textbf{Carl Kesselman} is the William H. Keck Professor of Engineering in the USC Viterbi School of Engineering and Professor in  Daniel J. Epstein Department of Industrial and Systems Engineering; 
a USC Information Sciences Institute Fellow, where he directs the Informatics Systems Research Division; and Director of the Center of Excellence for Discovery Informatics in the Michelson Center for Convergent Biosciences. 
\fi

\bibliographystyle{plain}
\bibliography{refs}

\end{document}